\documentclass[10pt,letterpaper,twocolumn]{article} %% two column, final layout
\usepackage{ol2}
\usepackage[draft,implicit=false]{hyperref}
\usepackage{amsmath}
\usepackage{graphicx}% Include figure files

\begin{document}

\twocolumn[
%\draft
\title{Purely nonlinear disorder-induced localizations and their parametric amplification}
\author{Viola Folli$^{1,2,4}$, Katia Gallo$^{3}$ and Claudio Conti$^{1,2}$}
\affiliation{
$^1$Department of Physics, University Sapienza, Piazzale Aldo Moro, 5, 00185, Rome (IT)\\
$^2$ISC-CNR, UOS Roma Sapienza, Piazzale Aldo Moro 5, 00185, Rome (IT)\\
$^3$Department of Applied Physics, KTH Royal Institute of Technology, Roslagstullbacken 21,
106 91 Stockholm, Sweden\\
$^4$Fondazione Istituto Italiano di Tecnologia (IIT),
Center for Life Nano Science Viale Regina Elena 291, 00161 Roma (IT)\\
}
\email{viola.folli@gmail.com}
\date{\today}
\begin{abstract}
We investigate spatial localization in a quadratic nonlinear medium in the presence of randomness. By means of numerical simulations and theoretical analyses we show that, in the down conversion regime, the transverse random modulation of the nonlinear susceptibility generates localizations of the fundamental wave that grow exponentially in propagation. The localization length is optically controlled by the pump intensity which determines the amplification rate. The results also apply to cubic nonlinearities.
\end{abstract}
]

\maketitle
\begin{figure}[t]
\includegraphics[width=\columnwidth]{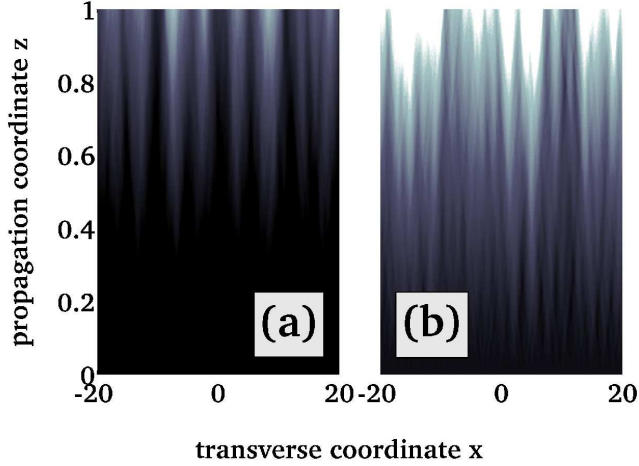}
\caption{(Color online) $A_1(x,z)^2$ after equation~(\ref{eqchi2}) for $\delta k=0.1$, $A_{1}(x,0)=0.01$, $d_0=1$ with $A_2(x,0)=A_{20}r(x)$ calculated for: $A_{20}=2$ (a) and $A_{20}=7$ (b). In the latter case, the localized states are strongly enhanced by the more effective disordered potential $V(x)=d_0A_{20}r(x)$.}
\label{fig1} \end{figure}
\begin{figure}[t]
\includegraphics[width=\columnwidth]{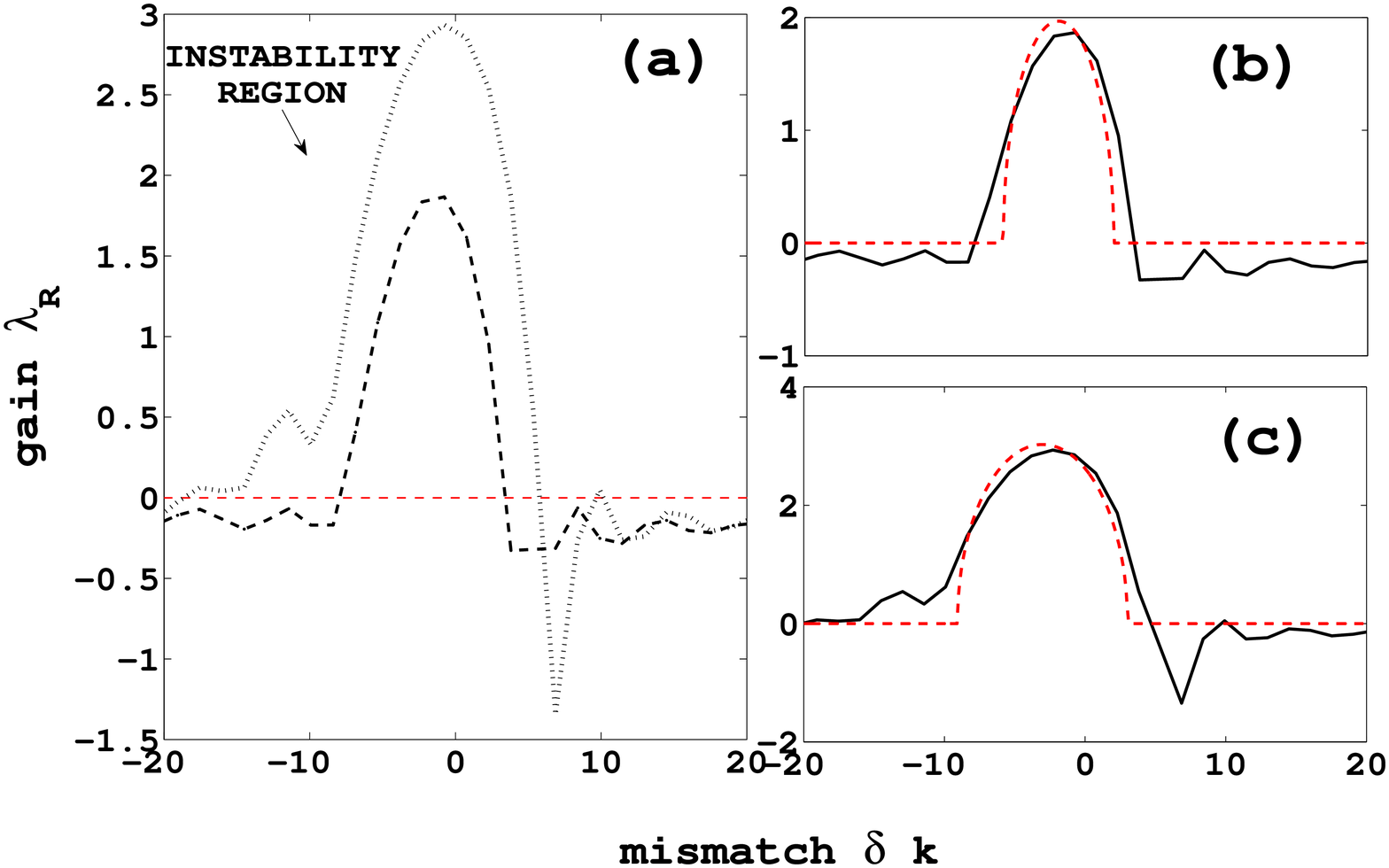}
\caption{(Color online) (a) Numerically calculated gain $\lambda_R$ after Eq.(\ref{eqchi2}) versus $\delta k$ for $A_{1}(x,0)=0.01$ with $A_{20}=2$ (\textit{dashed line}) and $A_{20}=3$ (\textit{dotted line}). The eigenvalues of the ground state are $E_0\simeq-1.2$ and $E_0\simeq-1.7$, respectively. The theoretical values of the instability region $\delta k_1=-2E_0$ and $\delta k_2=6E_0$ are in a good agreement with the simulations. Data are fitted by Eq.~\ref{lambda_R} with $V_{00}=2E_0$, $f^2(\delta k)=(3b-\delta k/2)(b+\delta k/2)$ (\textit{dashed lines}) for: $A_{20}=2$ (b) and $A_{20}=3$ (c). The parameters are: $b=-0.9\pm0.7$ (b) and $b=-1.5\pm0.3$ (c). Solid lines in (b), (c), represent numerical data.}
\label{fig2} \end{figure}
\begin{figure}[t]
\includegraphics[width=\columnwidth]{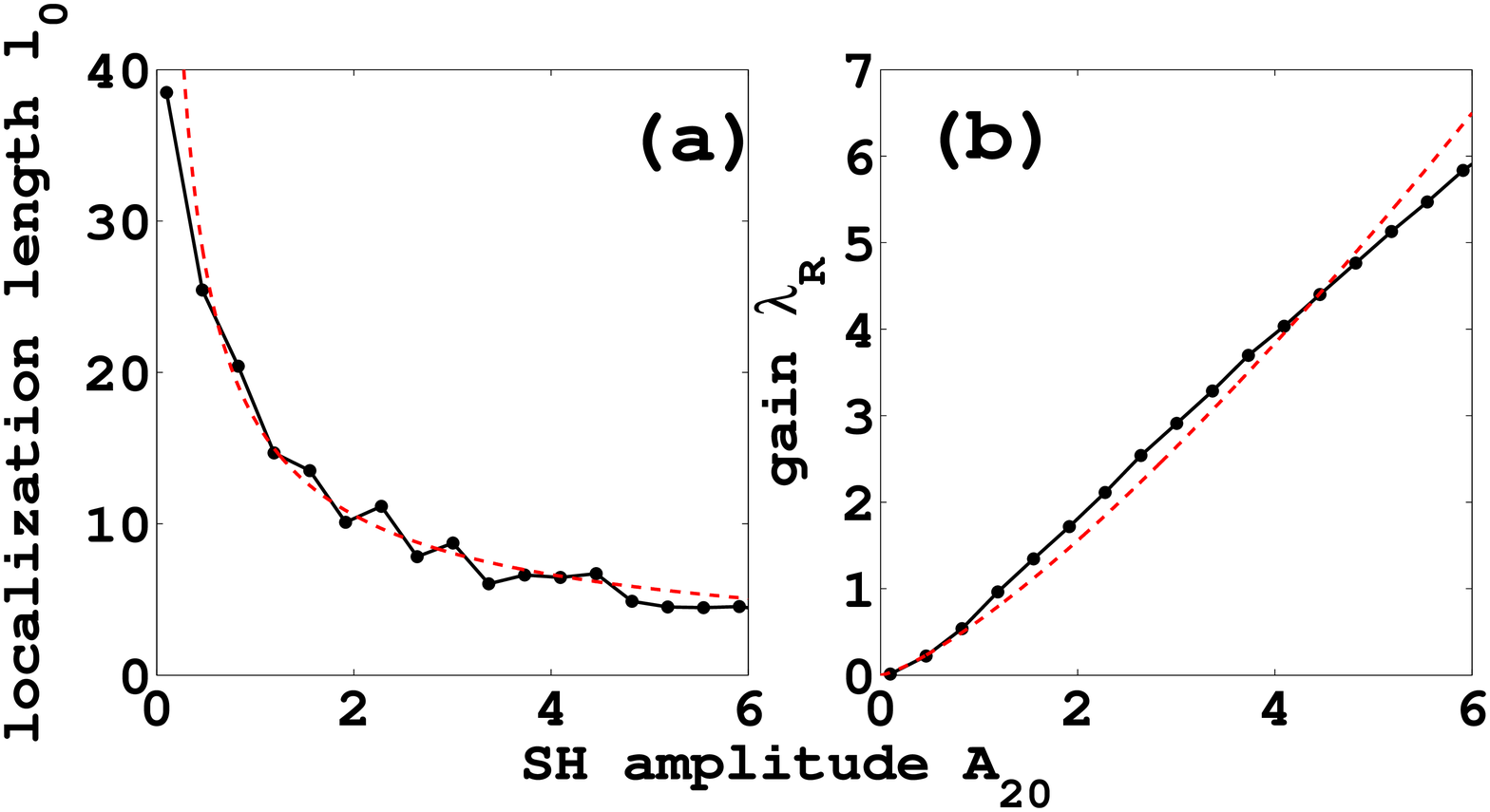}
\caption{(Color online) (a) Numerically found localization lengths for the ground state of Eq.~(\ref{anderson}) versus $SH$ amplitude $A_{20}$ (continuous line) with $A_{10}=0.01$, $\delta k=0.1$ and for $10$ realizations of disorder. Dashed line is the fit after Eq.~(\ref{scaling lambda}). (b) Gain $\lambda_R$ versus $A_{20}$ (continuous line) for $\delta k=0.1$ and $A_{10}=0.01$. Dashed line is the fit after Eq.~(\ref{scaling growth}).}
\label{fig3} \end{figure}

%% %%%%%%%%%%%%
\noindent Over the past few years, the interplay between nonlinearity and randomness generated relevant interest, specifically in the fields of light-matter interaction~\cite{Bodyfelt,Kivshar1990,Shadrivov2010,fishman2012nonlinear} and Bose-Einstein condensation~\cite{InguscioNature2008}. In all these studies, disorder is given by a randomly varying linear response. In the presence of a disordered potential, multiple scattering may yield localized states, the Anderson localizations (ALs)~\cite{anderson58,Wiersma97,segev2013anderson,maret2013}, which decay exponentially over a characteristic length $l$. Nonlinearity is always introduced as a process affecting observables like the degree of localization, or destroying the disorder-induced Anderson states. This kind of analysis has been reported by several authors, for example in non-resonant systems by the propagation of solitons in local and 
nonlocal media~\cite{folli2010frustrated,Sacha09,contisoli,folli2012anderson,kartashov08} or shock-waves \cite{ghofraniha2012} and in resonant systems by the propagation of Self-Induced Transparency (SIT) pulses~\cite{folli2011self}. Related phenomena are absence of equilibrium, glassy regimes and complexity~\cite{flach2009universal,Leuzzi09}.\\In this Letter, 
we analyze the dynamics of the existence of a kind of light localization with a purely nonlinear origin~\cite{conti2000energy}. We observe that unstable localized states arise in a linearly homogeneous medium with a random modulation of the nonlinear response. The predicted effect is general and occurs for various kinds of nonlinearity. We consider a quadratic nonlinearity $\chi^{(2)}$~\cite{menyuk1994solitary}, for which experiments can be envisaged by parametric down-conversion in nonlinear lattices~\cite{gallo:161113} with disordered quasi-phase matching (QPM)~\cite{kivsharspeditoviamail} and the case of cubic nonlinearity $\chi^{(3)}$~\cite{Garanovich2012}
attainable, for example, in Bose-Einstein condensation by modulating the nonlinear interaction through Feshbach resonance~\cite{modugno2010anderson}. We stress these are localizations only due to a random modulation of the nonlinear response; albeit they are unstable, they can be observed for short propagation distances (of the order of the diffraction length). 
We also show that random nonlinearity produces a parametric amplification that enhances the localized states during their evolution. In addition, at variance with the linear case, the nonlinear localization length is determined by the optical fluence, i.e. by the input pump beam for $\chi^{(2)}$ and $\chi^{(3)}$. The analysis presented in what follows concerns frequency degenerate down-conversion via $\chi^{(2)}$ nonlinearities, hence the optical pump will be denoted as the second harmonic ($SH$), while the degenerate outputs (signal=idler) as the fundamental frequency ($FF$). The main conclusions of the analysis remain valid in the non-degenerate case. Furthermore they can also be extended to $\chi^{(3)}$ processes.\\
%%%%%%%%%%%%%%%%%%%%%
\noindent We start from the normalized $\chi^{(2)}$ coupled-mode system~\cite{menyuk1994solitary,Conti02c,Stivala10} for slowly varying envelopes of continuous-wave light beams, propagating in a randomly distributed quadratic nonlinearity
\begin{equation}
 \label{eqchi2}
\begin{array}{l}
i\partial_z A_1+\partial_x^2A_1+d(x)A_1^*A_2=0\\
i\partial_z A_2+\displaystyle\frac{1}{2}\partial_x^2A_2+\delta k A_2+\displaystyle\frac{1}{2}d(x)A_1^2=0,\\
\end{array}
\end{equation}
where $\delta k =(2k_1-k_2)L_D$ is the wave-vector mismatch and $L_D=2k_1w_0^2$ is the diffraction length ($w_0$ is the beam waist); $d(x)$ is the normalized effective scaled second-order nonlinear coefficient, $x$ is the transverse coordinate in units of $w_0$ and $z$ is the propagation coordinate in units of $L_D$. $A_1$ and $A_2$ are respectively the amplitudes of the normalized fields at the fundamental frequency ($FF$) and second harmonic ($SH$); details can be found in~\cite{Conti02c}.\\
Randomness is introduced either by a Gaussian random nonlinear coefficient $d(x)$ or, in the considered undepleted pump approximation ($|A_2|>>|A_1|$),
by using a spatially modulated $SH$ pump beam $A_{2}(x,0)$.
In the former case, we consider a QPM profile $d(x)$ such that the correlation function of the disordered potential is $\langle d(x)d(x')\rangle=d_0^2\delta(x-x')$ and the brackets denote average over disorder realizations. Various quasi-phase matching ($QPM$) profiles can be considered and provide analogue results. We anticipate that, in both cases, the disorder strength will be proportional to $d_0 A_2(x,0)$, hence the disorder degree can be controlled by acting on the nonlinear coefficient or on the pump intensity.\\Writing the $SH$ and $FF$ fields respectively as $A_2(x,z)\rightarrow A_2(x,z)e^{i\delta k z}$ and $A_1(x,z)\rightarrow A_1(x,z)e^{i\delta k z/2}$ and assuming $A_2(x,z)$ to be real and slowly depending on $z$, from Eq.~(\ref{eqchi2}) we have for $FF$:
\begin{equation}
 \label{eqFF}
i\partial_zA_1+\partial_x^2A_1-V(x)A_1^*-\displaystyle\frac{\delta k}{2}A_1=0,
\end{equation}
where $V(x)=-d(x)A_2(x,0)$ represents the disordered potential. For the sake of simplicity we assume $A_2(x,0)$ as a real function, denoting the case in which the input SH beam is not spatially chirped, and
has only an amplitude modulation. We expand the solution $A_1$ of Eq.~(\ref{eqFF}) in term of the eigenstates of the potential function $V(x)$,  
\begin{equation}
\label{FF}
A_1(x,z)=\sum_n c_n(z)\varphi_n(x).
\end{equation}
The relevant disorder induced localizations $\varphi_n(x)$ are given by ($n=0,1,2,..$):
\begin{equation}
 \label{anderson}
-\frac{\partial^2\varphi_n}{\partial x^2}+V(x)\varphi_n(x)=E_n\varphi_n.  
\end{equation}
The states for $E_n<0$ are exponentially localized, such that $\varphi_n\simeq e^{-|x|/l_n}/\sqrt{l_n}$ with $l_n\simeq |E_n|^{-1/2}$~\cite{lif88}. We study their evolution in (\ref{eqFF}) by projecting over the state $\varphi_m(x)$ and obtaining the equation
\begin{equation}
\label{epsilon+-}
\begin{array}{l}
i\displaystyle\frac{d c_m}{dz}-c_m\Bigl(E_m+\displaystyle\frac{\delta k}{2}\Bigr)+\sum_n V_{mn}(c_n-c_n^*)=0
\end{array}
\end{equation}
being $\int \varphi_m(x)\varphi_n(x)dx=\delta_{mn}$ and $V_{mn}\equiv\int \varphi_m(x)\varphi_n(x)V(x)\simeq V_{mm}\delta_{mn}$, which is retaining one single term in the expansion, as the cross overlap among states is negligible. This approximation is valid for propagation distances smaller than the parametric gain length and is confirmed by the numerical results reported below.\\
The analysis of the exponentially diverging solutions of Eq.~(\ref{epsilon+-}) gives the stability properties of the nonlinear localizations, 
\begin{equation}
 \label{epsilon_lambda}
c_m(z)=\hat{c}_m e^{\lambda_R z},
\end{equation}
where $\lambda_R$ is the real-valued growth rate of the $m-th$ component of Eq.~(\ref{FF}). Substituting in Eq.~(\ref{epsilon+-}), we obtain 
\begin{equation}
\label{lambda_R}
\lambda_R^2=(2V_{mm}-E_m-\delta k/2)(E_m+\delta k/2).
\end{equation}
For $\delta k$ between $\delta k_1=-2E_m$ and $\delta k_2=-2E_m+4V_{mm}$, $\lambda_R^2$ is positive and, as it can be noted by Eq.~(\ref{epsilon_lambda}), the $m$th component of~(\ref{FF}) grows exponentially. The maximum growth rate $max(\lambda_R^{m})=-V_{mm}>0$ corresponds to $\delta k=2V_{mm}-2E_m<0$.\\As $|V_{mm}|$ is proportional to $A_{20}$, the instability range is determined by the strength of randomness, weighted by the input $SH$ amplitude, $A_{20}$. The overlap integral $V_{mm}=\int V(x) \varphi_m^2(x) dx\equiv\langle V\rangle$ can be related to the $m$th eigenvalue $E_m$: 
\begin{equation}
\int V(x) \varphi_m^2(x) dx=E_m\int \varphi_m^2 dx+\int\varphi_{mxx}\varphi_m dx,
\end{equation}
by using the averaged state, $\varphi(x)\simeq e^{-|x|/l}/\sqrt{l}$, we have
\begin{equation}
 \label{Ukk}
V_{mm}=E_m-\frac{1}{l_m^2}\simeq 2E_m<0,
\end{equation}
so that the instability growth-rate directly gives the localization eigenvalue and the localization length.\\
\noindent We point out the relation between the localization length and the rate of growth with the $SH$ amplitude. Being $\langle V(x)V(x')\rangle=V_0^2\delta(x-x')$ with $V_0=d_0A_{20}$, and letting $x=\tilde{x}/V_0^{2/3}$ and $E_n=\tilde{E}_nV_0^{4/3}$, we obtain the scaled equation $-\varphi_{\tilde{x}\tilde{x}}+r(\tilde{x})\varphi=\tilde{E}\varphi$ with $\langle r(\tilde{x})r(\tilde{x}')\rangle=\delta(\tilde{x}-\tilde{x}')$. The localization length for a state $\varphi$ can be hence written as
\begin{equation}
\label{scaling lambda}
l=\tilde{l}/V_0^{2/3}=\tilde{l}/(d_0A_{20})^{2/3},
\end{equation}
with $\tilde{l}$ the value for $V_0=1$.
The scaling law for the maximum growth-rate:
\begin{equation}
\label{scaling growth}
 max(\lambda_R^{m})=-V_{mm}=-2E_m=-2\tilde{E}_m(d_0A_{20})^{4/3}.
\end{equation}
By varying the $SH$ amplitude, we expect an increase of the growth-rate and an enhanced localization, as numerically verified below.\\
\noindent Numerically, we use a split-step beam propagation method (BPM) to integrate Eqs.~(\ref{eqchi2}), with the input condition $|A_1(x,z=0)|\ll|A_2(x,z=0)|$. We solve the eigenvalue problem~(\ref{anderson}) by a pseudo-spectral method for the $FF$ modes with potential $V(x)=d_0A_{20}r(x)$, and $r(x)$ a normalized Gaussian distribution, $\langle r(x)r(x')\rangle=\delta(x-x')$ with periodic boundary condition. The degree of localization depends on the strength of disorder $V_{0}=d_0A_{20}$. In Fig.~\ref{fig1} we show the $FF$ during propagation for two values of $A_{20}$. The evolution reveals the role of the $SH$ field as a disorder potential that induces localizations for the $FF$, a process that is more evident when increasing the amount of $SH$, i.e., the strength of randomness.
%%%%%%%%%%%%%%%FIGURE 1 %%%%%%%%%%%%%%%%%%%%%%
%%%%%%%%%%%%%%%%%%%%%%%%%%%%
In order to test the previous theoretical analysis, we use the ground-state of~(\ref{anderson}) as initial condition for the $FF$ and measure the growth-rate $\lambda_R$ for the localization during the evolution for several wave-vector mismatches. We fit by an exponential the evolution of the $FF$ peak at a fixed disorder realization. In Fig.~\ref{fig2}, we show $\lambda_R$ as a function of $\delta k$. The negative values of $\lambda_R$ in figure~\ref{fig2} are 
due to the unsuitability of the fit outside the instability $\delta k$ region. The found instability regions are in agreement with Eq.~(\ref{lambda_R}).
%%%%%%%%%%%%%%%FIGURE 2%%%%%%%%%%%%%%%%%%%%%%
%%%%%%%%%%%%%%%%%%%%%%%%%%%%%%%%%%%%%%%%%%%%%
In Fig.~\ref{fig3}, panel (a), we show the localization length of the ground state at $z=1$ versus $A_{20}$ and for $\delta k=0.1$ (in proximity of the maximum expected growth-rate). As the $SH$ amplitude increases, the strength of disorder grows and the states get more localized. Correspondingly, the localization becomes more unstable and the gain $\lambda_R$ increases (see Fig.~\ref{fig3}, panel (b)). The curves follow the scaling laws, Eqs.~(\ref{scaling lambda}) and (\ref{scaling growth}).
%%%%%%%%%%%%%%%FIGURE 3%%%%%%%%%%%%%%%%%%%%%%
%%%%%%%%%%%%%%%%%%%%%%%%%%%%%%%%%%%%%%%%%%%%%
To investigate the dynamics of the localized states, we consider an initial wave function $A_1(x,0)=A_{10}e^{-\beta x^2}$.
%%%%%%%%%%%%%%%FIGURE 4 %%%%%%%%%%%%%%%%%%%%%%
\begin{figure}[t]
\includegraphics[width=7cm]{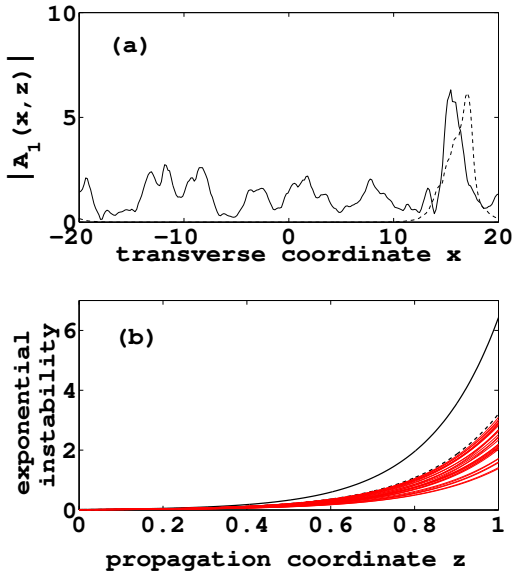}
\caption{(Color online) (a) $FF$ amplitude $|A_1(x,z)|$ at $z=1$ (continuous line) and the ground state arbitrarily scaled for: $A_{01}=0.01$, $\delta k=0.1$, $A_{20}=10$. (b) Evolution of the peak value of the $FF$ wave (continuous thick line) and coefficients $c_m(z)$, in Eq.~\ref{psit}, for: $m=0$ (dashed line) and $m>0$ (continuous thin lines).}
\label{fig4} \end{figure} 
%%%%%%%%%%%%%%%%%%%%%%%%%%%%%%%%%%%%%%%%%%%%%
For a flat input ($\beta=0$, $A_{10}=0.01$), we compare in Fig.~\ref{fig4}, panel (a), the profile of the $FF$ after propagation ($z=1$, continuous line) with the ground state of Eq.~(\ref{anderson}) (dashed line); the $FF$ gets more localized in proximity of the ground-state. Correspondingly, we expand $A_1(x,z)$ at any $z$ in terms of the eigenfunctions of Eq.~(\ref{anderson}):
\begin{equation} 
 \label{psit}
A_1(x,z)=\sum_m c_m(z)\varphi_m(x),
\end{equation}
the coefficients $c_m(z)$, found numerically and shown in figure~\ref{fig4}, panel (b), confirm the predicted exponential trend. The FF state evolves towards the ground-state and the $FF$ growth-rate (continuous thick line) is closer to $c_0(z)$ (dashed line).
%%%%%%%%%%%%%%%FIGURE 5 %%%%%%%%%%%%%%%%%%%%%%
\begin{figure}[t]
\includegraphics[width=8cm]{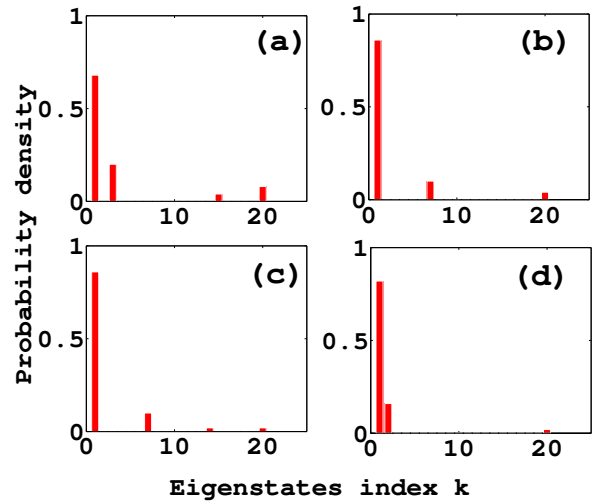}
\caption{(Color online)  Histogram of the minimum difference between the growth rate of $A_1(x,z)$ and $\varphi_k$, with $50$ realizations of disorder, $A_{01}=0.01$ and $\delta k=0.1$. (a) $A_{20}=1$, (b) $A_{20}=2$, (c) $A_{20}=3$, (d) $A_{20}=4$. }
\label{figure5} \end{figure} 
%%%%%%%%%%%%%%%%%%%%%%%%%%%%%%%%%%%%%%%%%%%%%
Similar dynamics occurs by taking $A_{1}(x,0)$ as a Gaussian state, $\beta=1$. Statistically, the state that has the nearest growth-rate to $A_{1}(x,z)$ is the mostly localized one, i.e., the ground state as shown by the histograms in Fig.~\ref{figure5}.\\We have obtained similar results in the case of Kerr media with a disordered cubic nonlinear term, $\chi(x)$, resulting in a randomly modulated nonlinear Schroedinger equation, $i\partial_z A+\partial_x^2A+\chi(x)|A|^2A=0$, where $A(x,z)$ is the scaled field amplitude. Repeating the analysis above, this system admits exponentially unstable localizations with a growth-rate of ${\lambda^m_R}^2=-E_m^2+\chi_{mm}^2$, where $E_m$ is the $m$-th eigenvalue and $\chi_{mm}=-\int dx \varphi_m^2(x)\chi(x)|A_0(x)|^2$, being $A_0(x)$ a slowly varying field envelope. The numerical investigations for Kerr media will be reported elsewhere.\\We stress that the predicted localizations are unstable, because subject to parametric $\chi^{(2)}$ or hyper-parametric $\chi^{(3)}$ amplification, however they can be observed for finite propagation. As an example, in the down-conversion regime, assuming $d_0\simeq1$ $pm/V$ with $w_0\sim10$ $\mu m$ and an effective transverse waveguide length of the order of $10$ $\mu m$, a peak power of $100$ $kW$ at $\lambda=1$ $\mu m$ results in a localization length of the order of $50$ $\mu m$. \\In conclusion, the parametric down-conversion of a light beam in a quadratic medium in conjunction with a random transverse modulation of the nonlinear susceptibility (or, equivalently, of the pump beam) brings about the generation of localized regions that grow exponentially with the intensity of the $SH$ field. In addition, we show that the localization length of these states can be controlled by the optical fluence, which also determines the strength of disorder. Our results also apply to other kind of nonlinearities, such as a randomly modulated Kerr nonlinearity, and can be extended to 2D and 3D cases and have implication in classical frequency down-
conversion devices and for quantum optical applications, such as parametric sources of entangled photon pairs. We remark that nonlinear amplification can be used to analyze the states induced by a given distribution of disorder and hence retrieve the properties of the latter, thus providing a characterization tool for the disorder that can associated to e.g. periodic poling in ferroelectric crystals or related approaches.

\end{document}